\documentclass[12pt]{iopart}
\usepackage{graphicx}
\usepackage{bm}
\usepackage{citesort}
\usepackage{iopams}

\newtheorem{theorem}{Theorem}[section]
\newtheorem{proposition}[theorem]{Proposition}

\renewcommand{\epsilon}{\varepsilon}

\newtheorem{Remark}[theorem]{Remark}

\newcommand{\PRB}{{\it Phys. Rev. B} }

\newcommand{\PRE}{{\it Phys. Rev. E} }

\newcommand{\APL}{{\it Appl. Phys. Lett.} }

\newcommand{\jpa}{{\it J. Phys. A: Math. Theor.} }

\begin{document}
\title[Dirichlet wave guide with two Neumann discs]{Spectral and localization properties of the Dirichlet wave guide with two concentric Neumann discs}
\author{H Najar}
\address{D\'epartemente de Math\'ematiques, ISMAI, Kairouan, Bd Assed Ibn Elfourat, 3100 Kairouan Tunisia}
\ead{hatem.najar@ipeim.rnu.tn}
\author{O Olendski$^{1,2,3}$} \address{$^1$Department of Physics, Jackson State University, 1400 Lynch St., Jackson, MS 39217 USA}
\address{$^2$Atomic and Molecular Engineering Laboratory, Belarusian State University, Skarina Av. 4, Minsk 220050 Belarus}
\address{\footnote{Present address}$^3$King Abdullah Institute for Nanotechnology, King Saud University, P.O. Box 2455, Riyadh 11451 Saudi Arabia}
\ead{oolendski@ksu.edu.sa}

\begin{abstract}
Bound states of the Hamiltonian describing a quantum particle living on three dimensional straight strip of width $d$ are investigated. We impose the Neumann boundary condition on the two concentric windows of the radii $a$ and $ b$ located on the opposite walls and the Dirichlet boundary condition on the remaining part of the boundary of the strip. We prove that such a system exhibits discrete eigenvalues below the essential spectrum for any $a,b>0$. When $a$  and $b$ tend to the infinity, the asymptotic of the eigenvalue is derived. A comparative analysis with the one-window case reveals that due to the additional possibility of the regulating energy spectrum the anticrossing structure builds up as a function of the inner radius with its sharpness increasing for the larger outer radius. Mathematical and physical interpretation of the obtained results is presented; namely, it is derived that the  anticrossings are accompanied by the drastic changes of the wave function localization. Parallels are drawn to the other structures exhibiting similar phenomena; in particular, it is proved that, contrary to the two-dimensional geometry, at the critical Neumann radii true bound states exist.
\end{abstract}
\vskip.7in

\noindent

\pacs{03.65.Ge, 03.65.Nk, 02.30.Tb, 73.22.Dj}

\maketitle

\section{Introduction}
\label{sec_1}

The task of finding eigenenergies $E_n$ and corresponding eigenfunctions $f_n({\bf r})$, $n=1,2,...$ of the Laplacian in the two- (2D) and three-dimensional (3D) domain $\Omega$ with mixed Dirichlet 
\begin{equation}\label{Dirichlet1}
\left.f_n({\bf r})\right|_{\partial\Omega_D}=0
\end{equation}
and Neumann 
\begin{equation}\label{Neumann1}
\left.{\bf n}{\bm\nabla}f_n({\bf r})\right|_{\partial\Omega_N}=0
\end{equation}
boundary conditions on its confining surface (for 3D) or line (for 2D) $\partial\Omega=\partial\Omega_D\cup\partial\Omega_N$ ($\bf n$ is a unit normal vector to $\partial\Omega$) \cite{Prange1,Borisov1,Dowker1,Borisov2,Wiersig1,Seeley1,Driscoll1,Levitin1,Borisov3,Borisov4,Jakobson1,Holcman1} is commonly referred to as Zaremba problem \cite{Zaremba1}. Apart from the purely mathematical interest, an analysis of such solutions is of a large practical significance as they describe miscellaneous physical systems. For example, the temperature $T$ of the solid ball floating in the icewater  obeys the Neumann condition on the part of the boundary which is in the air while the underwater section of the body imposes on $T$ the Dirichlet demand \cite{Dowker1}. Mixed boundary conditions were applied for the study of the spectral properties of the quantized barrier billiards \cite{Wiersig1} and of the ray splitting in a variety of physical situations \cite{Prange1}. The problem of the Neumann disc in the Dirichlet plane emerges naturally in electrostatics \cite{Jackson1}. In the limit of the vanishing Dirichlet part of the border the reciprocal of the first eigenvalue describes the mean first passage time of Brownian motion to $\partial\Omega_D$ \cite{Holcman1}. In cellular biology, the study of the diffusive motion of ions or molecules in neurobiological microstructures essentially employs the combination of these two types of the boundary coniditons on the different parts of the confinement \cite{Schuss1}.

One class of Zaremba geometries that recently received a lot of attention from mathematicians and physicists are 2D and 3D straight and bent quantum wave guides \cite{Exner1,Exner3,Bulla1,Exner6,Davies1,Dittrich1,Borisov5,Dittrich2,Borisov10,Olendski1,Borisov6,Borisov7,Borisov12,Johnson2,Popov1,Olendski2,Gortinskaya1,Olendski3,Trifanova1,Krejcirik1,Borisov13,Borisov8,Borisov9,Najar1,Borisov11,Assel1}. In particular, the conditions for the existence of the bound states and resonances in such classically unbound system were considered for the miscellaneous permutations of the Dirichlet and Neumann domains \cite{Dittrich1,Borisov6,Olendski2}. Bound states lying below the essential spectrum of the corresponding straight part were predicted to exist for the curved 2D channel if its inner and outer interfaces support the  Dirichlet and Neumann requirements, respectively, and not for the opposite configuration \cite{Dittrich2,Olendski1,Krejcirik1}. This was an extension of the previous theoretical studies of the existence of the bound states for the pure Dirichlet bent wave guide  \cite{Exner2,Schult1} that were confirmed experimentally  \cite{Carini1}. Magnetic field influence on the Dirichlet-Neumann structures was analyzed too \cite{Borisov13,Olendski2,Olendski3}. Also, for the 2D straight Dirichlet wave guide the existence of the bound state below the essential spectrum was predicted when the Neumann window is placed on its confining surface \cite{Exner1,Bulla1}. From practical point of view, such configuration can be realized in the form of the two window-coupled semiconductor channels of equal widths \cite{Exner1,Exner6} whose experimental creation and study has been made possible \cite{Hirayama1} due to the advances of the modern growth nanotechnologies. The number of the bound states increases with the window length $L$ and their energies are monotonically decreasing functions of $L$ \cite{Borisov10}. Recently, this result was extended to the case of the 3D spatial Dirichlet duct with circular Neumann disc \cite{Najar1} for which a proof of the bound state existence was confirmed, the number of discrete eigenvalues as a function of the disc radius $a$ was evaluated and their asymptotics for the large $a$ was given. As mentioned above, such Zaremba configuration is indispensable for the investigation of the electrostatic phenomena \cite{Jackson1}. Similar to the 2D case, it can be also considered as  the equal widths limit of the two 3D coupled Dirichlet ducts of, in general, different widths with the window in their common boundary \cite{Exner6,Borisov12}. Another motivation  stems from the phenomenological Ginzburg-Landau theory of superconductivity \cite{deGennes1} which states that the boundary condition for the order parameter $\Psi({\bf r})$ of the superconducting electrons reads
\begin{equation}\label{Robin1}
\left.{\bf n}{\bm\nabla}\Psi\right|_{\it\partial\Omega}=\left.\frac{1}{\Lambda}\Psi\right|_{\it\partial\Omega},
\end{equation}
$\bf n$ being an inward unit vector normal to the confining interface $\it\partial\Omega$. Extrapolation length (or de Gennes distance) $\Lambda$ is equal to infinity for the superconductor/dielectric boundary and tends to zero for the contact with ferromagnets \cite{deGennes1}. Thus, placing atop the superconductor the materials with these two limiting  extrapolation lnegths, one inevitably needs to deal with mixed boundary conditions.

In the present research, we discuss the case of the two concentric Neumann windows with, in general, different radii $a$ and $b$ on the opposite walls of the 3D straight wave guide.  A comparative analysis with the one Neumann disk reveals that the two-window geometry offers more possibilities for the varying spectral and localization properties by the additional channel of changing the second disc radius. Namely, its variation leads to the formation of the anticrossing structure of the energy spectrum as a function of smaller radius $b$ when the almost flat parts of the energy $b$-dependence are alternated by their sharp drops down to the lower level pushing it to occupy the next lying below neighbouring state, etc. Transitions  get sharper and the gaps between the anticrossing states decrease with increasing the outer radius  $a$. These avoided crossings are accompanied by the drastic changes of the localization of the wave function.

The rest of the paper is organized as follows. In section \ref{sec_2}, we define the model and recall some known results. In section \ref{sec_3}, we present the main result of this investigation followed by a discussion. Section \ref{sec_4} is devoted for numerical experiments, and concluding remarks are collected in section \ref{sec_5}.

\section{Model and Formulation}
\label{sec_2}
\begin{figure}
\centering
\includegraphics[width=0.95\columnwidth]{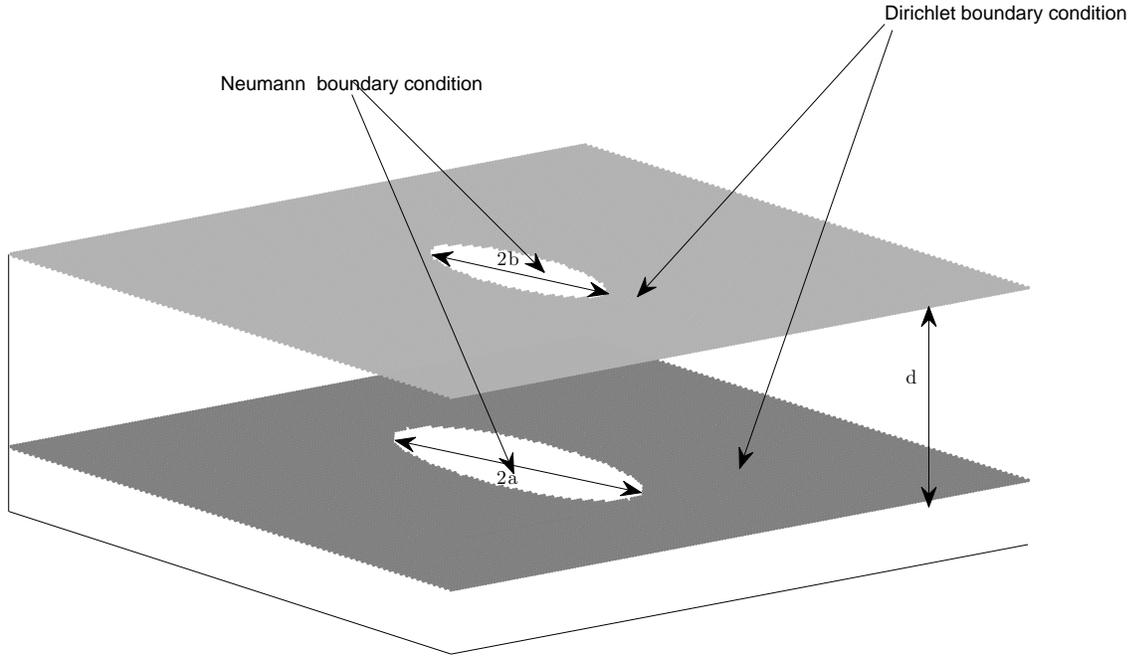}
\caption{\label{Fig1}
Dirichlet wave guide with two concentric Neumann disc windows on the opposite walls with (in general) different radii $a$ and $b$.}
\end{figure}

The system we are going to study is given in figure \ref{Fig1}. We consider a Schr\"odinger particle with mass $m_p$ whose motion is confined to a pair of parallel planes separated by the width $d$. For simplicity, we assume that they are placed at $z=0$ and $z=d$. We shall denote this configuration space by $\Omega$
\[
\Omega=\mathbb{R}^2\times [0,d].
\]

Let $\gamma_0(a)$ be a disc of radius $a$ with its center at $(0,0,0)$ and $\gamma_d(b)$ be a disc of radius $b$ centered at $(0,0,d)$, where
\begin{equation}
\gamma.(z)=\{(x,y,\cdot)\in \mathbb{R}^3;\ x^2+y^2\leq z^2\}.
\end{equation}
Without loss of generality we assume that $0\le b\le a$. We set $\Gamma=\partial \Omega\diagdown (\gamma_0(a)\cup \gamma_d(b))$. We consider Dirichlet boundary condition on $\Gamma$ and Neumann boundary condition in $\gamma_0(a)$ and $\gamma_d(b)$ what means that $\partial\Omega_D$ and $\partial\Omega_N$ from \eref{Dirichlet1} and \eref{Neumann1} take the following form: $\partial\Omega_D\equiv\Gamma$, $\partial\Omega_N\equiv\gamma_0(a)\cup \gamma_d(b)$.

\subsection{The Hamiltonian}\label{Hamiltonian1}
Let us define the self-adjoint operator on $L^{2}(\Omega )$ corresponding to the particle Hamiltonian $\hat{H}$. This is done by means of the quadratic forms. Namely, let $q_0$ be the quadratic form
\begin{equation}
\fl q_{0}(f,g)=\int_\Omega\overline{\nabla f}\cdot\nabla gd{\bf r},\ \mathrm{%
with\ domain}\ \mathcal{Q}(q_{0})=\{f\in H^{1}(\Omega );\ f\lceil\partial\Omega_D=0\},
\end{equation}
where $H^{1}(\Omega )=\{f\in L^{2}{(\Omega )}|\nabla f\in L^{2}(\Omega )\}$ is the standard Sobolev space and we denote by $f\lceil\partial\Omega_D$ the trace of the function $f$ on $\partial\Omega_D$. It follows that $q_{0}$ is a densely defined, symmetric, positive and closed quadratic form. We denote the unique self-adjoint operator associated with $q_{0}$ by $\hat{H}$ and its domain by $D(\Omega )$. It is the Hamiltonian describing our system. From \cite{Resi} (page 276), we infer that the domain $D(\Omega)$ of $\hat{H}$ is
\[
D(\Omega )=\Big\{f\in H^{1}(\Omega );\ -\Delta f\in L^{2}(\Omega
),f\lceil\partial\Omega_D=0,\frac{\partial f}{\partial z}\lceil\partial\Omega_N=0\Big\}
\]
and
\[
\hat{H}f=-\Delta f,\ \ \forall f\in D(\Omega ),
\]
where we have set $\hbar^2/(2m_p)\equiv 1$.

\subsection{Some known facts}\label{SomeKnownFacts1}

Let us start this subsection by recalling that in the particular case when $a=b=0$, we get $\hat{H}^{0}$, the Dirichlet Laplacian, and at $a=b=+\infty $ we get $\hat{H}^{\infty }$, the Neumann Laplacian. Since
\[
\hat{H}=\left(-\Delta_{\mathbb{R}^2}\right)\otimes I\oplus I\otimes\left(-\Delta_{\lbrack 0,d]}\right),
\mathrm{on}\ L^2(\mathbb{R}^{2})\otimes L^2([0,d]),
\]
(see \cite{Resi}) we get that the spectrum of $\hat{H}^0$ is $\left[\left(\frac{\pi}{d}\right)^{2},+\infty\right[$ and the spectrum of $\hat{H}^\infty$ is $[0,+\infty[$. Consequently, we have
\[
\left[\left(\frac{\pi}{d}\right)^2,+\infty\right[\subset\sigma
\left(\hat{H}\right)\subset\left[ 0,+\infty\right[ .
\]
This leads to a natural choice of the unit of energy as $\left(\pi/d\right)^2$ used below. In addition, if not stated otherwise, we will measure all distances in units of the wave guide width $d$, and all momenta, in units of $1/d$.

Using the property that the essential spectra are preserved under compact perturbation, we deduce that the essential spectrum of $\hat{H}$ for any finite $a$ and $b$ is
\[
\sigma _{ess}\left(\hat{H}\right)=\left[1,+\infty\right[.
\]
An immediate consequence is  that the discrete spectrum, if it exists, lies in $\left[0,1\right]$.

\subsection{Preliminary: Cylindrical coordinates}
\label{Cylindrical1}

As the system has a cylindrical symmetry, it is natural to consider the cylindrical coordinates system ${\bf r}=(r,\theta ,z)$. Indeed, we have that
\[
L^{2}(\Omega ,dxdydz)=L^{2}(]0,+\infty \lbrack \times \lbrack 0,2\pi \lbrack
\times \lbrack 0,1],rdrd\theta dz).
\]
We denote by $\langle ,\rangle_{\bf r}$ the scalar product in $L^{2}(\Omega ,dxdydz)=L^{2}(]0,+\infty\lbrack\times\lbrack 0,2\pi\lbrack\times \lbrack 0,1],rdrd\theta dz)$ given by
\[
\langle f,g\rangle_{\bf r}=\int_{]0,+\infty\lbrack\times\lbrack 0,2\pi\lbrack\times\lbrack 0,1]}\overline{f}grdrd\theta dz.
\]
The Laplacian operator is written as
\begin{equation}\label{Laplacian1}
\Delta_{r,\theta ,z}={\bm\nabla}_{\bf r}^2=\frac{1}{r}\frac{\partial}{\partial r}\left(r\frac{\partial}{\partial r}\right)+\frac{1}{r^2}\frac{\partial ^{2}}{\partial\theta^2}+\frac{\partial^2}{\partial z^2}.
\end{equation}
Therefore, the eigenvalue equation is given by
\begin{equation}\label{vp}
-\Delta_{r,\theta ,z}f(r,\theta ,z)=\pi^2Ef(r,\theta ,z).
\end{equation}
Since the operator is positive, we set $\pi^2E=k^2$. Equation~\eref{vp} is solved by separating the variables and considering the function $f$ as a product
\begin{equation}\label{FunctionF1}
f(r,\theta ,z)=R(r)\cdot\Theta(\theta)\cdot Z(z).
\end{equation}
Plugging the last expression into~\eref{vp}, one first separates $Z$ by putting all the $z$ dependence into one term so that $\frac{Z^{\prime \prime }}{Z}$ can only be constant taken, for convenience, as $-s^{2}$ with its value determined by the boundary conditions at $z=0$ and $z=1$. Second, we separate the term $\frac{\Theta "}{\Theta}$ which has all the $\theta$ dependence. Using the fact that the problem has an axial symmetry and the solution has to be $2\pi$ periodic and single-valued in $\theta$, we obtain that $\frac{\Theta "}{\Theta}$ should be a constant $-m^2$ for $m\in\mathbb{Z}$. Finally, we get the following equation for $R$
\begin{equation}\label{bes}
R^{\prime\prime }(r)+\frac{1}{r}R^{\prime }(r)+\left(k^{2}-s^{2}-\frac{m^{2}}{r^{2}}\right)R(r)=0.
\end{equation}
We notice that equation~\eref{bes} is the Bessel equation and its solutions could be expressed in terms of the Bessel functions \cite{abra,wat}. More explicit solutions could be given by considering boundary conditions; for example, for the Dirichlet requirements at the both walls one has $s_j^{DD}=(j+1)\pi$ while for the NN situation $s_j^{NN}=j\pi$, and for the mixed case $s_j^{ND}=(j+1/2)\pi$, $j\in\mathbb{N}$ (see Section \ref{sec_4} for more discussion).

\section{Analytical results}\label{sec_3}
Here we prove existence conditions and provide evaluations derived from the analytical consideration.
\begin{theorem}\label{th1}
The operator $\hat{H}$ has at least one isolated eigenvalue in $\left[0,1\right] $ for any nonzero $a$ and $b$.

Moreover, for $b$ big enough and $\lambda(a,b)$ 
being an eigenvalue of $\hat{H}$ less then $\displaystyle 1$,
there exist positive constants $C_a$ and $C_b$ such that
\begin{equation}\label{as1}
\lambda(a,b)\in\left(\frac{C_a}{a^2},\frac{C_b}{b^2}\right).
\end{equation}
\end{theorem}
\begin{Remark}
The first claim of Theorem \ref{th1} is valid for more general shape of bounded surface $\mathcal{S}$ with Neumann boundary condition, not necessarily a disc; it suffices that the surface contains a disc of radius $a>0$.
\end{Remark}
For the more general shape $\mathcal{S}$, using discs of radii $a$ and $a'$ such that
\begin{equation}
\gamma_0(a)\subset\mathcal{S}\subset\gamma_0(a'),
\end{equation}
and a comparison argument, one gets the localization of the discrete spectrum (see also Refs. \cite{Exner6,Borisov12}). 

\textbf{Proof.} For  the proof of the the first claim one may mimic the argument given in \cite{Najar1} and adjusted for the case of the two windows; however, much simpler and elegant way is to use the fact that the Neumann window is a negative perturbation \cite{Exner6}. Thus, if the one Neumann window creates the bound state \cite{Najar1}, the insertion of the second one just pushes it lower. $\blacksquare$

The proof of the second claim is based on the bracketing argument. Let us split $L^2(\Omega,rdrd\theta dz)$ as follows: $L^2(\Omega,rdrd\theta dz)=L^2(\Omega_{a,b}^{-},rdrd\theta dz)\oplus L^{2}(\Omega_{a,b}^{+},rdrd\theta dz)$, with
\begin{eqnarray*}
\Omega_{a,b}^{-}&=&\left\{(r,\theta ,z)\in \left[
0,b+(a-b)(1-z)\right]\times\lbrack 0,2\pi
\lbrack\times\lbrack 0,1]\right\},\\
\Omega_{a,b}^{+}&=&\Omega\backslash\Omega_{a,b}^{-}.
\end{eqnarray*}
Therefore,
\[
\hat{H}_{a,b}^{-,N}\oplus\hat{H}_{a,b}^{+,N}\leq\hat{H}\leq\hat{H}_{a,b}^{-,D}\oplus\hat{H}_{a,b}^{+,D}.
\]
Here, we index by $D$ and $N$ depending on the boundary conditions considered on the surface $r=b+(a-b)(1-z)$. The min-max principle leads to
\[
\sigma _{ess}\left(\hat{H}\right)=\sigma _{ess}\left(\hat{H}_{a,b}^{+,N}\right)=\sigma_{ess}\left(\hat{H}_{a,b}^{+,D}\right)=\left[1,+\infty \right[.
\]
Hence, if $\hat{H}_{a,b}^{-,D}$ exhibits a discrete spectrum below $\displaystyle 1$, then $\hat{H}$ does as well. We mention that this is not a necessary condition. If we denote by $\lambda_{j}\left(\hat{H}_{a,b}^{-,D}\right),\lambda_{j}\left(\hat{H}_{a,b}^{-,N}\right)$ and $\lambda_{j}\left(\hat{H}\right)$ the $j$-th eigenvalue of $\hat{H}_{a,b}^{-,D}$, $\hat{H}_{a,b}^{-,N}$ and $\hat{H}$, respectively, then, again, the minimax principle yields the following
\begin{equation}
\lambda _{j}\left(\hat{H}_{a,b}^{-,N}\right)\leq\lambda _{j}\left(\hat{H}\right)\leq\lambda_{j}\left(\hat{H}_{a,b}^{-,D}\right)\label{es1}
\end{equation}
and for $j\geq 2$
\begin{equation}\label{Sequence1}
\lambda _{j-1}\left(\hat{H}_{a,b}^{-,D}\right)\leq\lambda _{j}\left(\hat{H}\right)\leq\lambda_{j}\left(\hat{H}_{a,b}^{-,D}\right).
\end{equation}
As the computation of the eigenvalue of a frustum is not an easy task, let us remark that
\begin{equation}\label{Sequence2}
\lambda_j\left(\hat{H}_{a,a}^{-,D}\right)\leq\lambda_j\left(\hat{H}_{a,b}^{-,D}\right)\leq\lambda_j\left(\hat{H}_{b,b}^{-,D}\right)
\end{equation}
(the same chain of inequalities is true for the corresponding Neumann Hamiltonians $\hat{H}^{-,N}$, too). 
Then, from equations (\ref{Sequence1}) and (\ref{Sequence2}) it follows:
\begin{equation}\label{Sequence3}
\lambda _{j-1}\left(\hat{H}_{a,a}^{-,D}\right)\leq\lambda _{j}\left(\hat{H}\right)\leq\lambda_{j}\left(\hat{H}_{b,b}^{-,D}\right).
\end{equation}

The Hamiltonian $\hat{H}_{\xi,\xi}^{-,D}$ has a sequence of eigenvalues \cite{abra,wat} given by
\[
\lambda _{lmn}(\xi)=\left(l\pi\right) ^2+\left(\frac{x_{|m|n}}{\xi}\right)^2,
\]
where $x_{|m|n}$ is the $n$th positive zero of Bessel function of the order $|m|$ \cite{abra,wat} and index $j$ amalgamates the three  quantum numbers: transverse $l$, radial $n$ and azimuthal $m$ ones, $j\equiv(l,m,n)$. The condition
\begin{equation}\label{gaza}
\lambda _{lmn}\le 1
\end{equation}
yields that $l=0$, so we get $\lambda_{0mn}(b)=\left(x_{|m|n}/b\right)^2$ and $\lambda_{0m'n'}(a)=\left(x_{|m'|n'}/a\right)^2$ with primed $n$ and $m$ referring to the index $j-1$ in (\ref{Sequence3}). Accordingly, for any eigenvalue $\lambda(a,b)$ of the Hamiltonian $\hat{H}$, there exist $m,n,m',n' \in \mathbb{N}$ such that
\begin{equation}\label{as2}
\frac{x_{|m'|n'}^2}{a^2}\leq\lambda(a,b)\leq\frac{x_{|m|n}^2}{b^2}.
\end{equation}
When $b$ is big enough, we get the result. $\blacksquare$

The above derivation shows that the coefficients $C_a$ and $C_b$ in equation (\ref{as1}) are, actually, the squares of the zeros of the Bessel functions.
\begin{Remark}
As it is shown in the next section, for the large enough diameters $2b$ many bound states exist which are characterized by the azimuthal $m$ and radial $n$ quantum numbers. For each of them the estimate (\ref{as1}) holds; however, the magnitude of the smaller radius above which this estimate becomes true is different for the different states $(n,m)$. Equation (\ref{as2}) and the properties of the zeros of the Bessel function \cite{abra,wat} manifest that estimate (\ref{as1}) for the higher lying states is achieved at the larger $b$.
\end{Remark}

Next statement describes the case when the new bound state just emerges from the continuum what means that its energy is equal to the Dirichlet fundamental propagation threshold. We assume that the inner radius is either fixed in the interval $0\le b<a$ or is equal to its outer counterpart $a$.
\begin{proposition}\label{Proposition1}
When the radius  $a$ is  equal to a critical value $a_l$ at which a new bound state emerges from the continuum, equation \eref{vp} with $E=1$ has a solution $f_l^{(0)}(r,\theta,z)$, unique to a nonzero multiplicative constant which at infinity behaves like
\begin{equation}\label{FunctionAsymptotics}
\fl f_l^{(0)}(r,\theta,z)=\frac{e^{im\theta}}{\sqrt{2\pi}}\left[\frac{\sqrt{2}\sin\pi z}{r^{|m|}}+\beta_l\frac{e^{-\pi\sqrt{3}r}}{\sqrt{r}}\sin 2\pi z+{\cal O}\left(\frac{e^{-\pi\sqrt{8}r}}{\sqrt{r}}\right)\right],\quad r\rightarrow\infty
\end{equation}
with some constants $\beta_l$. Here, the two quantum numbers $n$ and $m$ are compacted into the single index $l$: $l\equiv(n,m)$.
\end{proposition}

A proof of this statement will be given in the next section.

Note that the derived asymptotics from \eref{FunctionAsymptotics} drastically differs from its quasi-one-dimensional couterpart \cite{Exner1,Exner3,Borisov10,Borisov6}; namely, if, in the latter case, the state at $E=1$ is {\it always} a resonance, for the 3D geometry considered in this paper this is true for the azimuthal quantum number $m=0$ only, and for $|m|\ge 2$ one has true bound states with square integrable functions $f_l^{\left(0\right)}\left({\bf r}\right)$. We point out that the function of the state with $|m|=1$ vanishes at infinity, however, it is not square integrable over the whole $r-\theta$ plane; accordingly, one can consider it as a resonance too. In other words, a behaviour of the states lying on the border between the essential and discrete spectra crucially depends on the dimensionality of the problem. 

\section{Numerical computations: mode-matching method}\label{sec_4}

This section is devoted to some numerical computations. We analyse a dependence of the eigenenergies $E$ and coresponding eigenfunctions $f$ on the radii $a$ and $b$ in the whole range of their variation. We also compare the obtained results with the one-window case \cite{Najar1}.

\subsection{Two identical Neumann windows}\label{Numeric1}

Consider first the case of the equal radii on the upper and lower walls of the duct. The eigenvalue equation~\eref{vp} uses the Laplacian provided in \eref{Laplacian1}. A general solution is written in equation \eref{FunctionF1}. Since for our problem a cylindrical symmetry is conserved, one has $\Theta(\theta)=\frac{1}{\sqrt{2\pi}}e^{im\theta}$. A total solution can be written:

for one Neumann window:
\numparts\label{SolutionOneWindow}
\begin{eqnarray}\label{SolutionOneWindow1}
\fl f_<(r,\theta ,z)=\frac{1}{\sqrt{2\pi}}e^{im\theta}\sum_{j=0}^\infty B_j^{|m|}J_{|m|}\left(\pi\sqrt{E-\left(j+1/2\right)^2}r\right)\chi_j^{ND}(z), \ r\le a\\
\label{SolutionOneWindow2}
\fl f_>(r,\theta ,z)=\frac{1}{\sqrt{2\pi}}e^{im\theta}\sum_{j=0}^\infty A_j^{|m|}K_{|m|}\left(\pi\sqrt{\left(j+1\right)^2-E}r\right)\chi_j^{DD}(z), \quad r\ge a;
\end{eqnarray}
\endnumparts

for the two identical concentric Neumann windows of the radius $a$ on the opposite walls:
\begin{equation}\label{SolutionTwoIdenticalWindows1}
\fl f_<(r,\theta ,z)=\frac{1}{\sqrt{2\pi}}e^{im\theta}\sum_{j=0}^\infty B_j^{|m|}J_{|m|}\left(\pi\sqrt{E-j^2}r\right)\chi_j^{NN}(z), \ r\le a
\end{equation}
and for $r\ge a$ it coincides with \eref{SolutionOneWindow2}. Here, $J_m(r)$ is Bessel function of the first kind with $K_m(r)$ being its modified counterpart of the second kind \cite{abra,wat}, coefficients $A_j^{|m|}$ and $B_j^{|m|}$ present a relative contribution of the corresponding component into the total wave function, and the orthonormalized functions $\chi_j^{DD}(z)$, $\chi_j^{ND}(z)$ and $\chi_j^{NN}(z)$ describing a transverse  motion in the Dirichlet-Dirichlet, Dirichlet-Neumann and Neumann-Neumann strips, respectively, read:
\numparts\label{TransverseFunctions}
\begin{eqnarray}
\label{TransverseDD}
\chi_j^{DD}\left(z\right)&=&\sqrt{2}\sin\left(j+1\right)\pi z\\
\label{TransverseND}
\chi_j^{ND}\left(z\right)&=&\sqrt{2}\cos\left(j+1/2\right)\pi z\\
\label{TransverseNN}
\chi_j^{NN}\left(z\right)&=&\left\{\begin{array}{c} 1,\ j=0 \\
                                                  \sqrt{2}\cos j\pi z,\ j\ne 0.
                                                  \end{array}
\right.
\end{eqnarray}
\endnumparts
Their corresponding eigenvalues are, respectively, $E_j^{DD}=(j+1)^2$, $E_j^{ND}=(j+1/2)^2$ and $E_j^{NN}=j^2$. Note that for $j\ne 0$ the Bessel functions $J_{|m|}\left(\pi\sqrt{E-\left(j+1/2\right)^2}r\right)$ in \eref{SolutionOneWindow1} and $J_{|m|}\left(\pi\sqrt{E-j^2}r\right)$ in \eref{SolutionTwoIdenticalWindows1} transform into their modified counterparts $I_{|m|}\left(\pi\sqrt{\left(j+1/2\right)^2-E}r\right)$ and $I_{|m|}\left(\pi\sqrt{j^2-E}r\right)$ \cite{abra,wat}, respectively.

Physically, the emergence of the bound state for at least one nonzero radius is due to the fact that the introduction of one or two Neumann windows mixes longitudinal and transverse motions in the wave guide what results in the splitting off of the unbound level from the lowest subband down below the fundamental propagation threshold and its corresponding transformation into the localized state with square-integrable wave function vanishing at infinity. Mathematically, this mixing is reflected in the fact that the summation index $j$ in equations 
(19)
and \eref{SolutionTwoIdenticalWindows1} enters each component of both transverse $Z(z)$ as well as radial $R(r)$ parts of the wave function binding them together. Note that this binding is different inside and outside the disc. In a sense, a Neumann perturbation of the Dirichlet duct presents a shelter where the Schr\"{o}dinger particle can dwell with its momentum smaller than the cutoff momentum of the fundamental propagation threshold. Similar to the bend in curved Dirichlet wave guide \cite{Sprung1}, one can draw parallels with the creation by the obstacle of  an attractive quantum well that binds the particle; namely, the wave can still be propagating inside the circle but vanishing outside: as equations \eref{SolutionOneWindow1} and \eref{SolutionTwoIdenticalWindows1} demonstrate, there is one propagating along the radial direction component [the term with $j=0$ when the difference $E-\left(j+1/2\right)^2$ (for one window) or $E-j^2$ (for the two discs) is positive] while outside the discs the wave function possesses exponentially vanishing contributions only.

Matching function $f$ and its radial derivative at $r=a$, one gets:
\begin{equation}\label{matrix1}
\sum_{j'=0}^\infty Q_{jj'}B_{j'}^{|m|}=0
\end{equation}
with the following matrix elements of the infinite square matrix ${\bf Q}$:
\begin{eqnarray}
Q_{jj'}&=&\left[\sqrt{\frac{E-\left(j'+1/2\right)^2}{\left(j+1\right)^2-E}} \frac{J'_{|m|}\left(\pi\sqrt{E-\left(j'+1/2\right)^2}a\right)}{K'_{|m|}\left(\pi\sqrt{\left(j+1\right)^2-E}a\right)}\right.\nonumber\\
\label{matrixQ}
&-&\left.\frac{J_{|m|}\left(\pi\sqrt{E-\left(j'+1/2\right)^2}a\right)}{K_{|m|}\left(\pi\sqrt{\left(j+1\right)^2-E}a\right)}\right]P_{jj'}^{(i)},
\end{eqnarray}
where the prime denotes a derivative of the Bessel function with respect to its argument and the superscript index $i=1(2)$ corresponds to one(two) Neumann window(s). The infinite square matrices ${\bf P}^{(1)}$ and ${\bf P}^{(2)}$ present the quantitative measure of the coupling between different modes of the DD and ND [for ${\bf P}^{(1)}$] or DD and NN [for ${\bf P}^{(2)}$] channels:
\begin{equation}\label{matrixP1}
P^{(1)}_{jj'}=\int_0^1\chi_j^{DD}(z)\chi_{j'}^{ND}(z)dz=\frac{2}{\pi}\frac{j+1}{(j+1)^2-(j'+1/2)^2}
\end{equation}
\begin{eqnarray}
P^{(2)}_{jj'}&=&\int_0^1\chi_j^{DD}(z)\chi_{j'}^{NN}(z)dz\nonumber\\
\label{matrixP2}
&=&\left\{
\begin{array}{cc} \frac{2}{\pi}\left[(-1)^{j+j'}+1\right]\frac{j+1}{(j+1)^2-{j'}^2}, &j'\ne 0,j'\ne j+1\\
0,& j'\ne 0,j'=j+1\\
\frac{\sqrt{2}}{\pi(j+1)}\left[(-1)^{j}+1\right], & j'=0.
\end{array}
\right.
\end{eqnarray}

Nontrivial solution of system \eref{matrix1} exists only when the determinant of the matrix ${\bf Q}$ is an identical zero:
\begin{equation}\label{eigenEquation1}
\det||{\bf Q}||=0.
\end{equation}
This is an equation for determining the bound-state energies $E$ for the different radii $a$. 
\begin{figure}
\centering
\includegraphics[width=0.95\columnwidth]{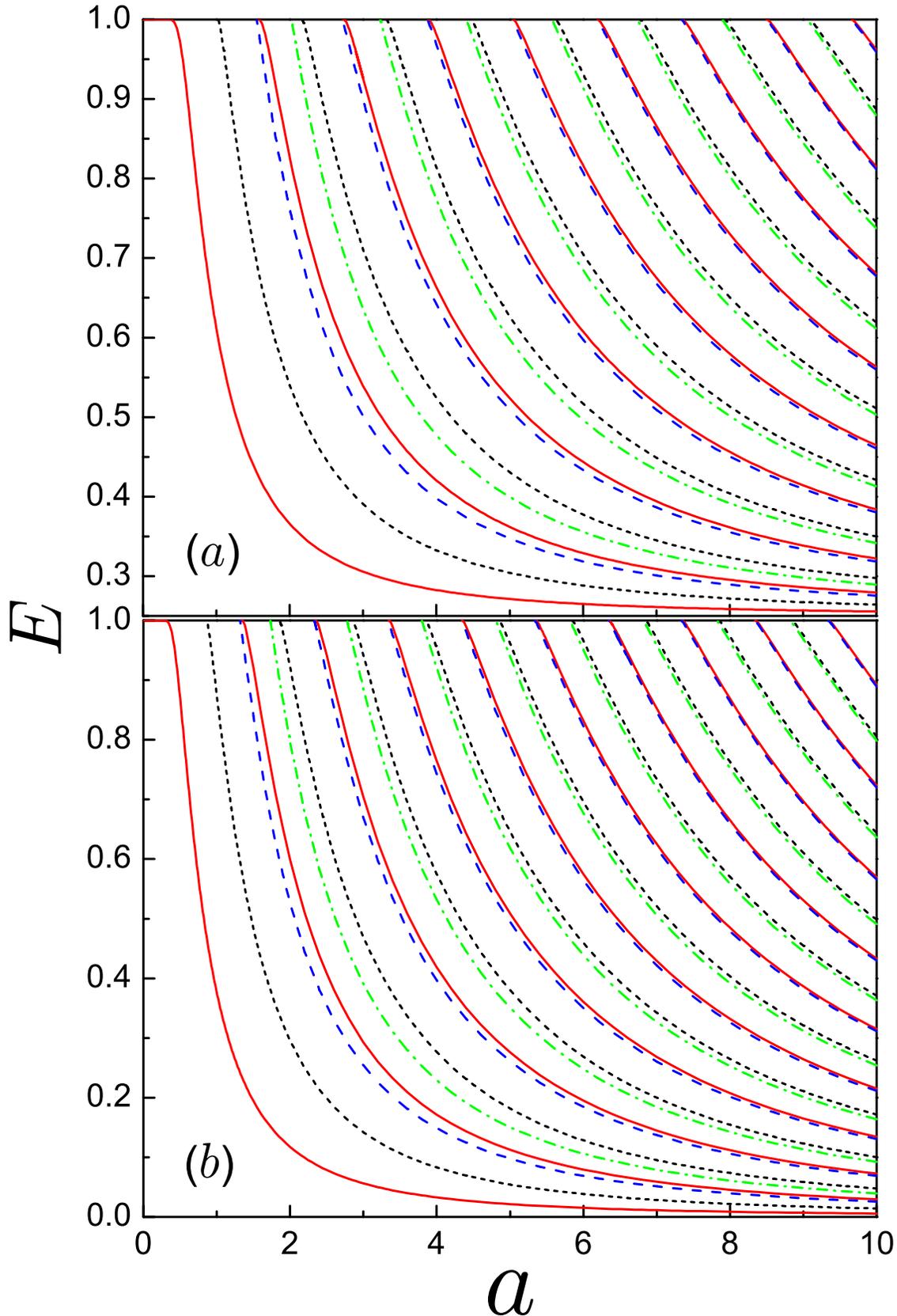}
\caption{\label{Fig2}
Bound-state energies $E$ as a function of the radius $a$ for (a) one Neumann window and (b) two concentric Neumann discs on the opposite sides of the wave guide where solid lines are for the states with the azimuthal quantum number $m=0$, dotted lines - for the states with $|m|=1$, dashed lines - for $|m|=2$, and dash-dotted curves - for $|m|=3$. Note different energy scales in panels (a) and (b).}
\end{figure}

From \eref{matrix1} the coefficients $B_j^{|m|}$ can be determined and, next, from the matching conditions, the values of $A_j^{|m|}$ are calculated too. Normalization condition
\begin{equation}\label{normalization1}
\int_0^{2\pi}d\theta\int_0^1dz\int_0^\infty drr\left|f\left(r,\theta,z\right)\right|^2=1
\end{equation}
allows one to fully construct the function $f$. Note that the radial integrals in \eref{normalization1} with the function $f$ given by equations
(19)
and \eref{SolutionTwoIdenticalWindows1} are calculated analytically \cite{Gradshteyn1,Prudnikov1} in terms of the Bessel functions themself.

\begin{figure}
\centering
\includegraphics[width=0.95\columnwidth]{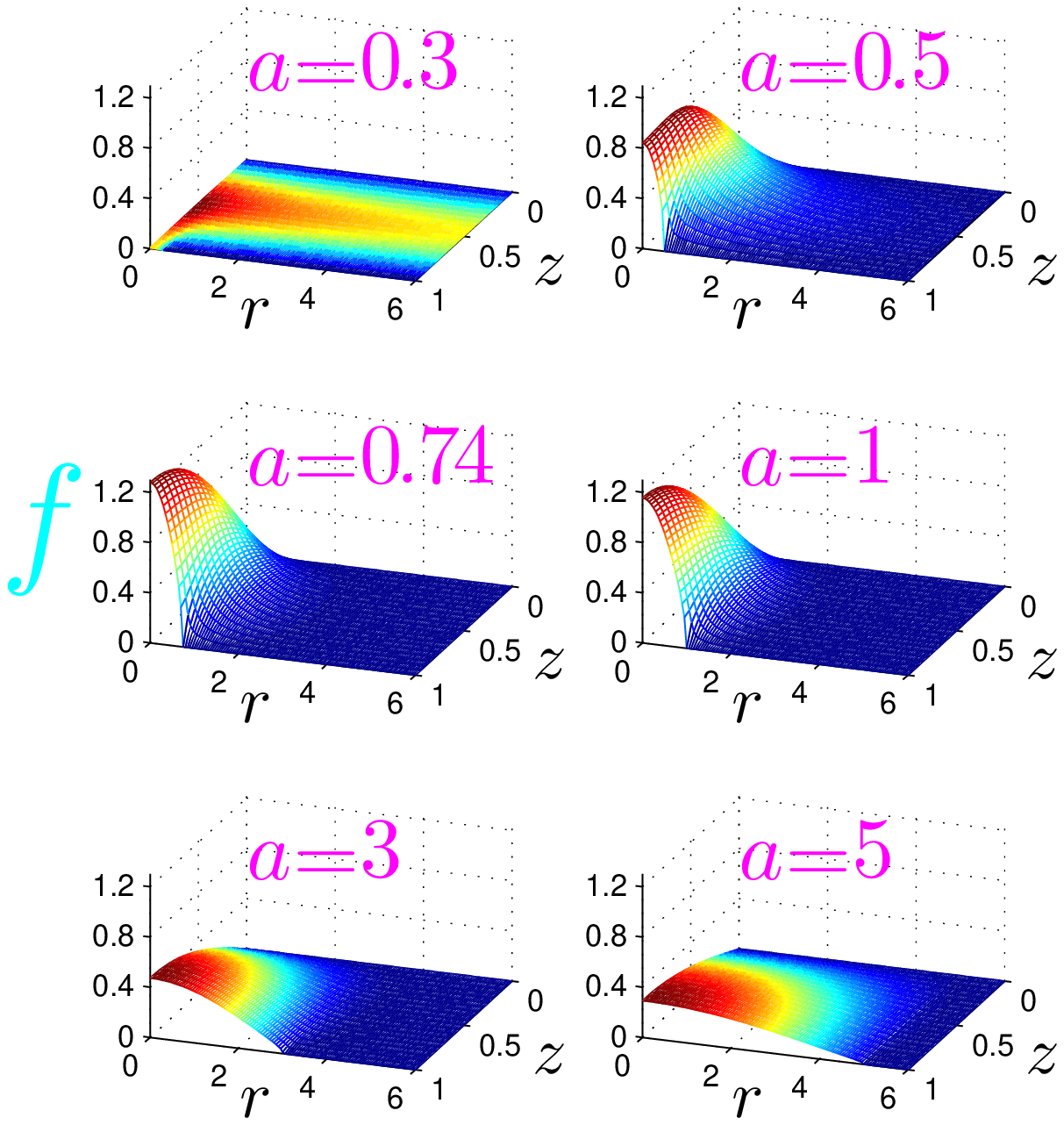}
\caption{\label{Fig3}
Ground bound-state wave function $f$ in terms of $r$ and $z$ for one Neumann window and several values of the radius $a$.}
\end{figure}

Fig. \ref{Fig2} depicts bound-state energies as a function of the normalized radius $a$. Both configurations show a lot of similarities. As discussed in \cite{Najar1} for one and earlier in this paper for the two identical Neumann windows, at least one bound state exists for the arbitrarily small radius $a$: in both cases this is the state with the azimuthal quantum number $m=0$. For quite small radii its energy lies very close to the fundamental propagation threshold of the uniform Dirichlet wave guide $E_0^{DD}=1$.  It was conjectured \cite{Exner6} that for the one window the energy of this single state at the small radius  depends on  $a$ as 
\begin{equation}\label{Evaluation1}
E(a)=1-\exp\left(-\frac{c}{a^3}\right), \quad a\rightarrow 0.
\end{equation}
Our results show that the constant $c$ is in the range $0.44\le c \le 0.45$. Large numerical errors for the energies close to the threshold precluded more accurate evaluation.

 Further growth of $a$ forces the bound-state energy to decrease, and at large radii it tends from above to the fundamental propagation threshold of the Neumann-Dirichlet $E_0^{ND}=1/4$ (for one window, ref. \cite{Najar1}) or the uniform Neumann $E_0^{NN}=0$ (for two windows, see above) channels. Second bound state possesses quantum number $|m|=1$ and emerges from the continuum at $a\sim 1.027$ for one window and $a\sim 0.866$ for the two Neumann discs. This difference is not surprising since the two Neumann regions present a stronger perturbation to the particle motion in the uniform Dirichlet wave guide as compared to the one window; accordingly, in the former case the second bound state is formed at the smaller radius. This is true for all other states (but the first one) as well; for example, the third bound state with $|m|=2$ emerges for the one disc at $a\sim 1.549$ and for the two Neumann windows - at $a\sim 1.319$. As our results show, in both cases the next bound level is another state with azimuthal quantum number $m=0$ preceding in this way its counterpart with $|m|=3$. Similar to the ground state, energies of the higher lying levels decrease when the radius increases with the difference between them diminishing what means the increase of the corresponding density of states (number of states per unit energy) ${\cal N}(E)$ until at $a=\infty$ one arrives at the continuous spectrum of the ND (one window) or NN (two windows) channel with the diverging density ${\cal N}(E)$.

\begin{figure}
\centering
\includegraphics[width=0.95\columnwidth]{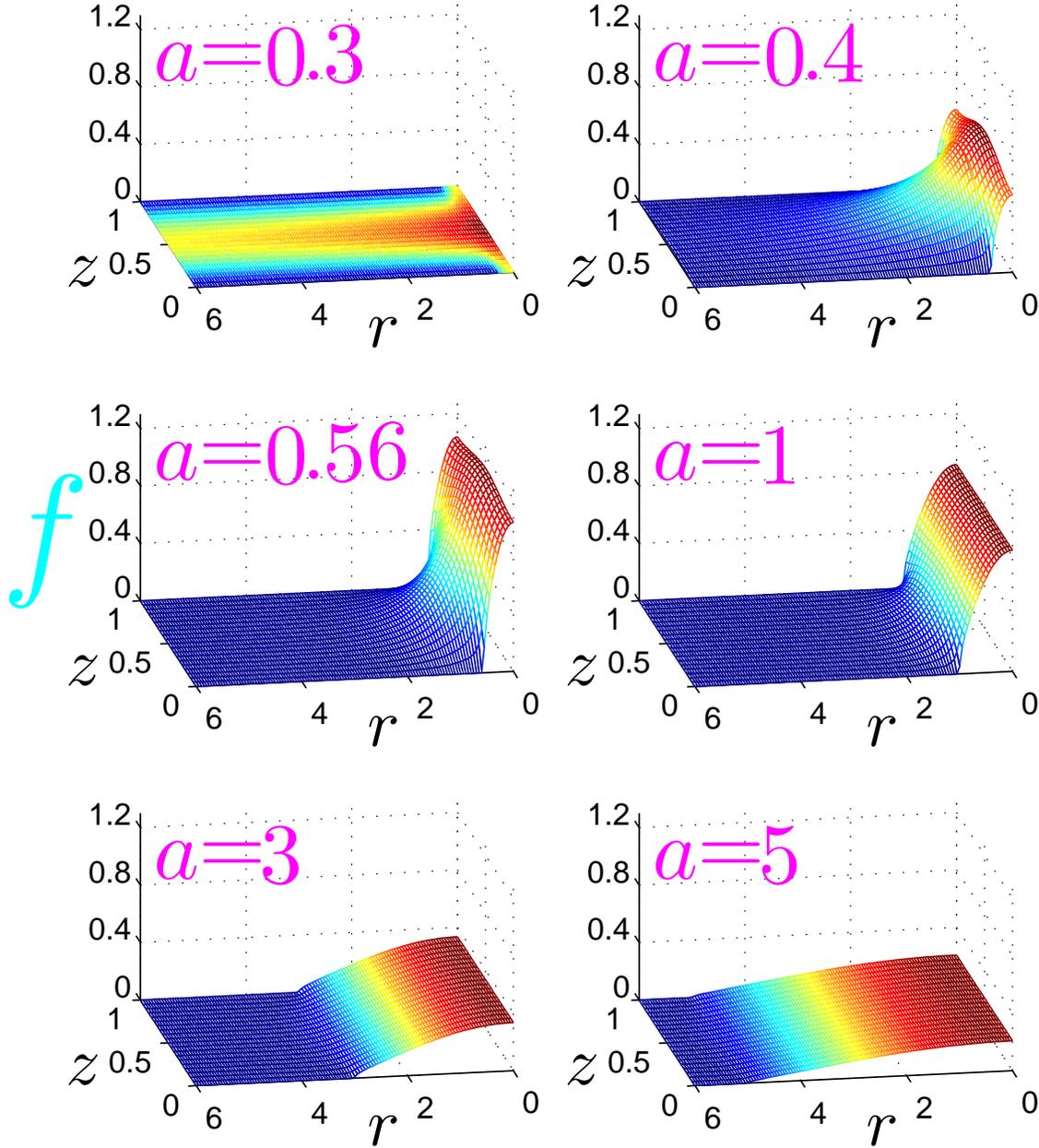}
\caption{\label{Fig4}
The same as in figure \ref{Fig3} but for the two concentric equal Neumann windows on the opposite walls.}
\end{figure}

Figures \ref{Fig3} and \ref{Fig4} depict the lowest bound-state wave function $f(r,z)$ for one and two windows, respectively, at several Neumann radii $a$. Since the lowest level possesses azimuthal quantum number $m=0$, its wave function is $\theta$-independent. It is seen that for the small radii, say, $a=0.3$ in figures \ref{Fig3} and \ref{Fig4}, the wave function is almost flat spreading far away from the Neumann area. This can be easily understood by means of the mentioned above analogy with the quantum well \cite{Landau1}; namely, for the small radius $a$ the effective positive potential created by the disc, only barely binds the particle with its wave function relatively slowly vanishing at infinity. Increasing Neumann radius strengthens the attractiveness of the well; accordingly, as the panels for $a=0.4$ (figure \ref{Fig4}) and $a=0.5$ (figure \ref{Fig3}) exhibit, the function $f$ is stronger localized inside the disc with the magnitude of its maximum growing. In both configurations the maximum is achieved at the centre of the disc, $r=0$. As our calculations show, for one window this extremum reaches it maximal value of $f_{max}\sim 1.28$ at $a\sim 0.74$ and for the two discs - at $a\sim 0.56$ with $f_{max}\sim 1.12$ (see corresponding panels in figures \ref{Fig3} and \ref{Fig4}). Subsequent increase of the Neumann area makes the potential well wider and wider  with the wave function $f$ being almost uniformly distributed inside it. Its radial slope  diminishes, as a transition from the situation with $a=1$ to the one with $a=3$ and, next, to $a=5$ in figures \ref{Fig3} and \ref{Fig4} vividly demonstrates. Such a behaviour can be easily explained implementing again the potential well model \cite{Landau1}. Introduction of the one Neumann disc breaks the wave guide symmetry with respect to the plane $z=1/2$. This is reflected in the corresponding asymmetry of the wave function which inside the disc, for the fixed radius $r<a$, reaches the maximum at $z=0$ and continuously decreases to zero at $z=1$. On the contrary, second identical Neumann window restores this symmetry with the function $f$ reflecting it in figure \ref{Fig4}.
\begin{figure}
\centering
\includegraphics[width=0.95\columnwidth]{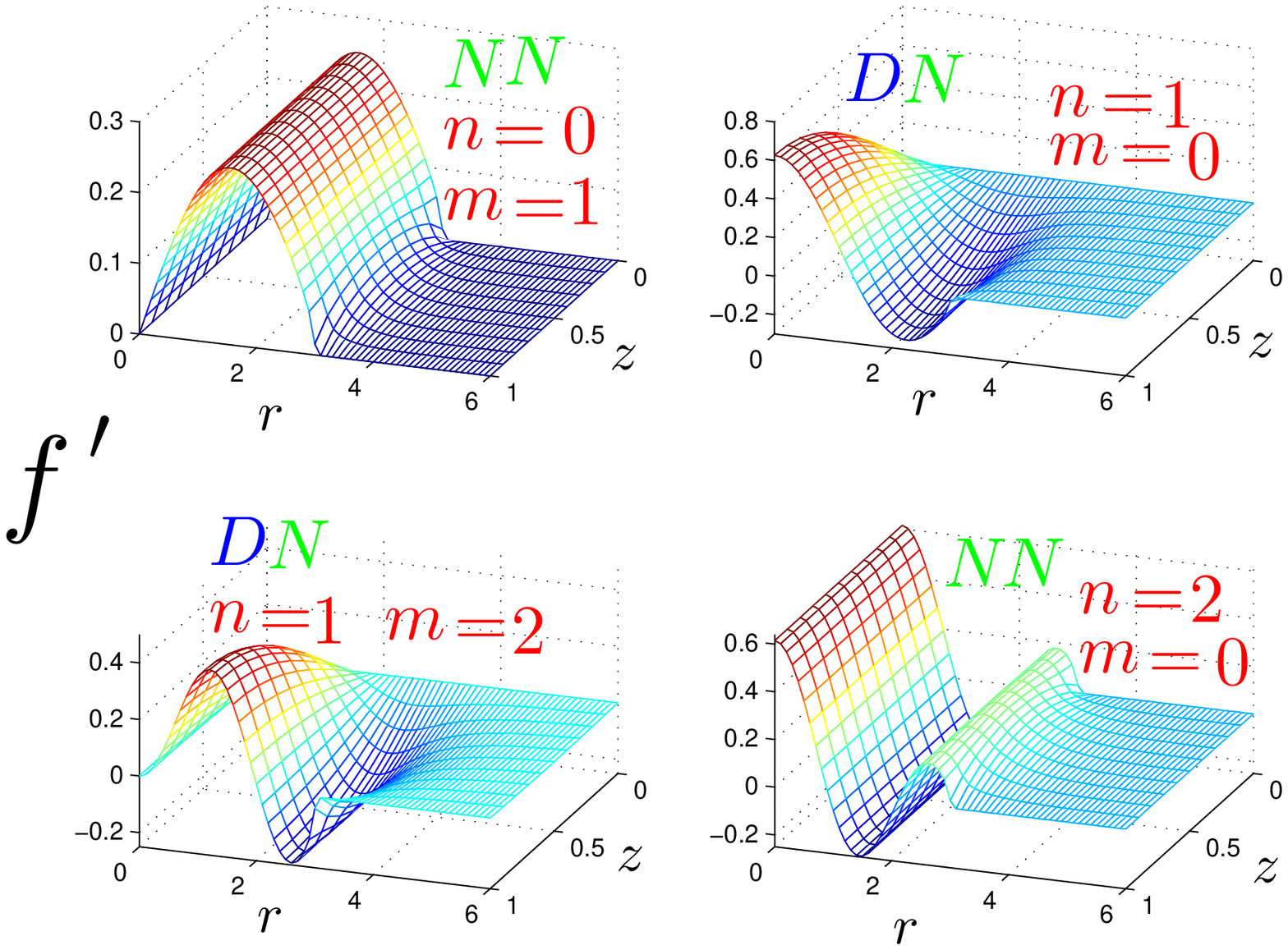}
\caption{\label{Fig5} Functions $f'\equiv e^{-im\theta}f$ in terms of $z$ and $r$ for the Neumann radius $a=3$ and several quantum numbers $n$ and $m$. Two capital characters denote a corresponding type of the boundary conditions on the opposite wave guide walls inside the circle $r\le a$. Note different $f'$-axis scales for each of the panels.}
\end{figure}

It was already mentioned in the Introduction that bound states are characterized by the two quantum numbers: principal $n$ and azimuthal $m$. To exemplify this, we plot in figure \ref{Fig5} functions $f'\equiv\exp(-im\theta)f$ for the fixed radius $a=3$ and several quantum numbers $(n,m)$ for both configurations of the boundary conditions. In addition to the discussed above features, one can notice that the state with the principal number $n$ has $n+1$ maxima on $r$ axis at the fixed $z$. For the nonzero angular momentum the wave function is zero at $r=0$ while for $m=0$ it achieves here the largest maximum. All these properties are in accordance with the general rules of quantum mechanics for the radially symmetric systems \cite{Landau1}. Note also that no any radial maxima and minima are observable in the semi infinite DD region: all of them are located in the NN and/or DN part, and the function $f$ exponentially decreases to zero as the radius swipes the $r$ axis to the infinity.

\subsection{Two different Neumann discs: anticrossings and their evolution}
Next, consider the case of the two concentric Neumann discs with different radii $a$ and $b$ on the opposite walls of the otherwise uniform Dirichlet channel. Only the region $b/a\le 1$ is under investigation since, as stated above, we assume that $b\le a$, $b\geq 0$. Here, one needs to consider three spatial regions: first, for $r\le b$, when the transverse motion is confined by the two Neumann plates; second, for $b\le r\le a$ when the motion in the $z$ direction is governed by the Neumann requirement on the one side and the Dirichlet demand on the opposite walls; and, the last, the region of $r>a$ with the pure Dirichlet transverse boundary conditions. Total solution of \eref{vp} for the first region is described by equation \eref{SolutionTwoIdenticalWindows1}, for the region of $r\ge a$ it is written as equation \eref{SolutionOneWindow2}, and for the ring $b\le r\le a$ one has:
\begin{eqnarray}
f(r,\theta ,z)&=&\frac{1}{\sqrt{2\pi}}e^{im\theta}\sum_{j=0}^\infty \left[C_j^{|m|}J_{|m|}\left(\pi\sqrt{E-\left(j+1/2\right)^2}r\right)\right.\nonumber\\
\label{SolutionRing1}
&+&\left. D_j^{|m|}Y_{|m|}\left(\pi\sqrt{E-\left(j+1/2\right)^2}r\right)\right]\chi_j^{ND}(z),
\end{eqnarray}
where the Bessel functions $J_{|m|}\left(\pi\sqrt{E-\left(j+1/2\right)^2}r\right)$ and $Y_{|m|}\left(\pi\sqrt{E-\left(j+1/2\right)^2}r\right)$  for $E<\left(j+1/2\right)^2$ transform into the modified Bessel $I_{|m|}\left(\pi\sqrt{\left(j+1/2\right)^2-E}r\right)$ and MacDonald $K_{|m|}\left(\pi\sqrt{\left(j+1/2\right)^2-E}r\right)$ functions, respectively, and unknown coefficients $C_j^{|m|}$ and $D_j^{|m|}$ are to be found from the matching of the function $f$ and its radial derivative at the circles $r=b$ and $r=a$. This matching leads again to the eigenvalue equation of the form of \eref{eigenEquation1}, where, however, in the present case the infinite matrix $\bf Q$ takes the block form:
\begin{equation}
{\bf Q}=\left[\begin{array}{cc}{\bf Q}^{(1)} &{\bf Q}^{(2)}\\
{\bf Q}^{(3)} &{\bf Q}^{(4)}
 \end{array}\right].
\end{equation}
\begin{figure}
\centering
\includegraphics[width=0.95\columnwidth]{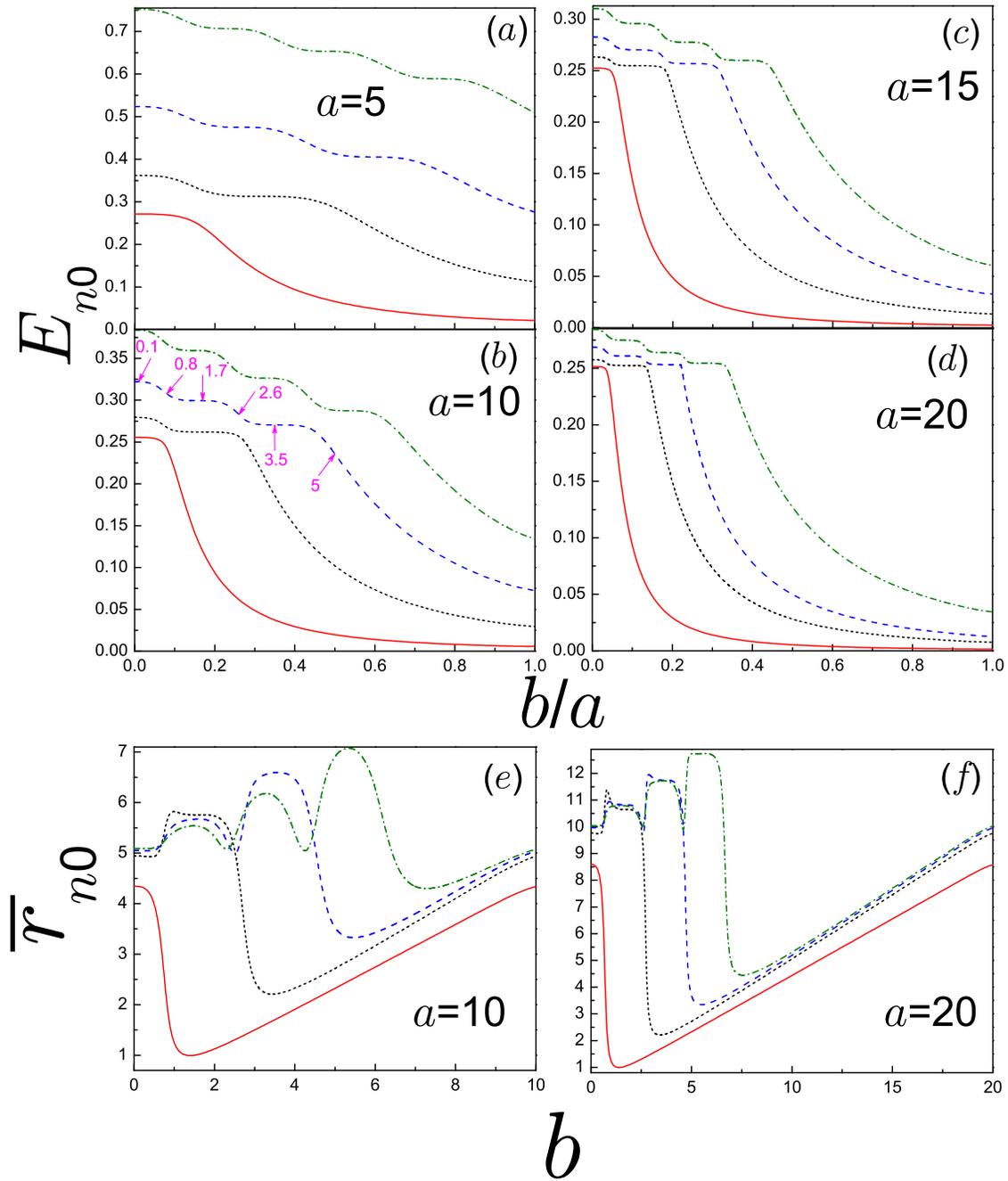}
\caption{\label{Fig6} (a)-(d) Energies $E_{n0}$ as functions of the ratio $b/a$ for several values of the outer radius $a$ where the panel (a) is for $a=5$, (b) - for $a=10$, (c) - for $a=15$, and (d) - for $a=20$. Solid lines denote the ground state ($n=0$), dotted lines depict the energies of the first excited level ($n=1$), dashed lines are for $n=2$, and dash-dotted lines  - for $n=3$. Note different energy scales for each of the figures. Arrows in panel (b) show energies for which the corresponding functions $f$ are plotted in figure \ref{Fig7}. Numbers near the arrows denote corresponding inner radius $b$. Panels (e) and (f) depict the mean radius $\overline{r}_{n0}$ as a function of the inner radius $b$ for $a=10$ [panel (e)] and $a=20$ [panel (f)]. The same convention as in panels (a)-(d) is used. Different $\overline{r}$ and $b$ scales are used in each of the panels.}
\end{figure}
Here, the elements of the infinite matrices ${\bf Q}^{(i)}$, $i=1,2,3,4,$ can be derived in a way similar to the previous subsection; in particular, in addition to the matrices ${\bf P}^{(1)}$ and ${\bf P}^{(2)}$, they also contain the matrix ${\bf P}^{(3)}$ describing the coupling between the pure Neumann transverse motion and its Neumann-Dirichlet counterpart:
\begin{equation}\label{matrixP3}
P^{(3)}_{jj'}=\int_0^1\chi_j^{ND}(z)\chi_{j'}^{NN}(z)dz=\left\{
\begin{array}{cc} (-1)^{j+j'}\frac{2}{\pi}\frac{j'+1/2}{(j'+1/2)^2-j^2}, &j\ne 0\\
(-1)^{j'}\frac{\sqrt{2}}{\pi}\frac{1}{j'+1/2}, & j=0.
\end{array}
\right.
\end{equation}
Energies $E_{n0}$ are plotted in panels (a)-(d) of figure \ref{Fig6} as a function of the radii ratio for several fixed outer radius $a$. Similar pictures are obtained for the nonzero azimuthal numbers $m$ too. As expected, for the vanishing inner radius, $b\rightarrow 0$, the energies approach those of the one Neumann window case \cite{Najar1} while for the $b\rightarrow a$ one recovers the two equal discs geometry discussed in the previous subsection. Note quite large slope $\partial E/\partial\left(b/a\right)$ of the lowest curves depicting the case of the large outer radii $a\ge 10$ [panels (b)-(d)]. The point is that here at both limiting parameters of  the inner radius the ground-state energy lies very close to the corresponding fundamental threshold: $E_{00}^{ND}=1/2$ for $b=0$ and  $E_{00}^{NN}=0$ for $b=a$. Accordingly, a change of the inner radius forces the particle to undergo transitions from the one situation to the other one and, since this energy interval is large compared to the smaller $a$, one observes rather steep energy descent on $b/a$ axis.

\begin{figure}
\centering
\includegraphics[width=0.95\columnwidth]{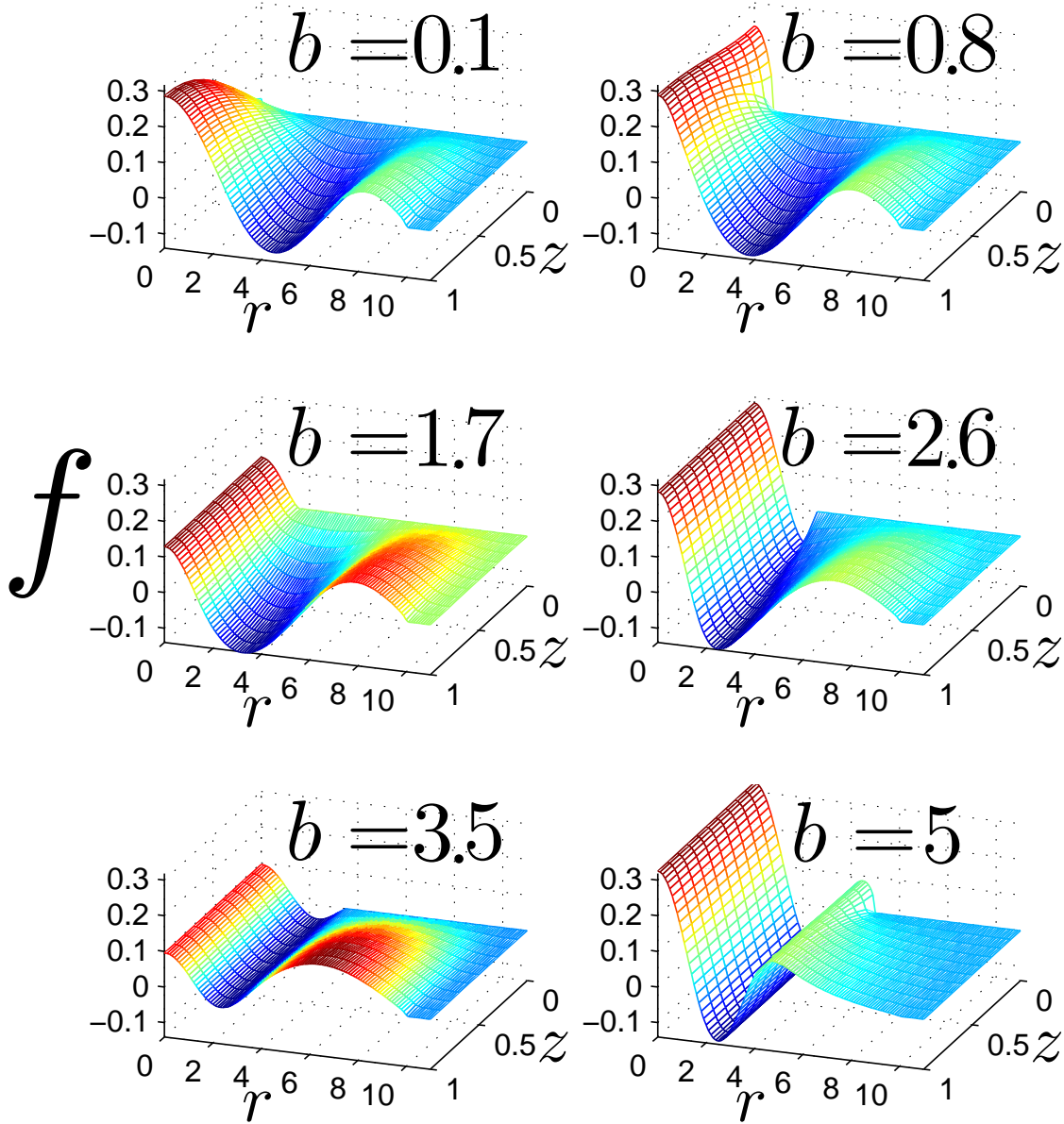}
\caption{\label{Fig7} Wave function $f_{2,0}$ in terms of $r$ and $z$ for the outer Neumann radius $a=10$ and
several inner radii $b$. Corresponding energies $E_{2,0}$ are shown by the arrows in panel (b) of figure \ref{Fig6}.}
\end{figure}

Another remarkable feature of the energy spectrum is emergence and evolution of the avoided crossings between the levels with the same azimuthal number $m$ and different principal numbers $n$. A hint of the anticrossing formation is already seen for the moderate outer radius $a=5$. As this radius increases, the interval between the levels at $b=0$ diminishes, as it was discussed in the previous subsection and reference \cite{Najar1}. For the moderate inner radius $b$ its change almost does not affect the energies. This is exemplified in panel (b) of figure \ref{Fig6} where the arrows show the energies at which the functions $f$ are plotted in figure \ref{Fig7} for the second excited level, $n=2$. It is seen that for the $n$th level the energy stays almost constant with $b$ growing until the expanding NN disc absorbs enough of the first maximum to alter the energy $E$. Further growth of $b$ and its subsequent approach to the first radial node drops the energy of the level to its lower counterpart and significantly distorts the NN maximum. For example, for the case of $b=0.8$ in panel (b) of figure \ref{Fig6} the corresponding part of figure \ref{Fig7} shows the conspicuous alteration of the first extremum with the first node located at $r\sim 2$. As soon as the expanding inner NN part of the wave guide approaches the next extremum of the function $f$, the energy $E_{nm}$ becomes strongly $b$-dependent. The rapid decrease of the energy in this narrow inner radius range is accompanied again by the strong modification of the wave function $f$. For example, it is seen from the comparative analysis of these two figures that for $b\sim 1.7$ the radial node of the function $f$ is close to the boundary between the NN inner disk and neighbouring ND annulus of the width $a-b$. As a result, we observe an almost flat plateau on the $E$-$b$ dependence. As the second extremum of the function $f$ is approached by the expanding NN disc, it gets strongly distorted what is clearly demonstrated by the corresponding panel for $b=2.6$ of figure \ref{Fig7}. This distortion of the wave function is accompanied by the simultaneous decrease of the corresponding energy $E_{nm}$ until it reaches the plateau of the neighbouring lower lying level with smaller $n$ pushing it downwards. This means that the function extremum fully penetrated into the NN disc; accordingly, the new flat part of the energy spectrum is formed when the boundary between the NN and ND geometries approximately coincides with the node of the function [see arrow for $b=3.5$ in panel (b) of figure \ref{Fig6} and the corresponding plot in figure \ref{Fig7}]. The gap between the $n$th and $(n-1)$th levels at the point of their closest encounter depends on the outer radius decreasing as $a$ increases. The sharpness of the transition between the plateaus increases for the larger outer radius too. Similar decrease of the anticrossing gap was predicted to exist for the other structures, for example, for the $N$ $\delta$-potentials with the distance between them growing \cite{Exner7}.  Such alternating plateau structure with the rapid change between them in the energy spectrum continues until all extrema of the function $f$ are absorbed by the expanding  NN disc after which the energy monotonically decreases to its value at $b=a$ [see arrow for $b=5$ in panel (b) of figure \ref{Fig6} and corresponding function plot in figure \ref{Fig7}]. Accordingly, each $n$th level has $(n+1)$ plateaus. The same structure as the one presented in figures \ref{Fig6} and \ref{Fig7} is observed for the nonzero azimuthal numbers $m$ too where, again, a flat plateau on the $E-b$ dependence corresponds to the NN-ND boundary being located in the vicinity of the node of the function $f$ when the change of the inner radius has a small influence on the eigenvalues of the corresponding Schr\"{o}dinger equation.

To understand these phenomena better, it is instructive to investigate the mean radius $\overline{r}$ according to
\begin{equation}\label{MeanRho}
\overline{r}_{nm}=\langle\Psi_{nm}|r|\Psi_{nm}\rangle.
\end{equation}
Its dependence on the inner radius $b$ is shown in panels (e) and (f) of figure \ref{Fig6} for $a=10$ and $a=20$, respectively. It is seen that each plateau in the energy spectrum is accompanied by the same flat part of the $\overline{r}$ dependence on the inner radius. Contrary to the energy spectrum, the transition between the plateaus is not smooth; namely, the increasing $b$ at the edge of one plateau initially pushes the wave function closer to the origin, as panels for $b=0.8$ and $b=2.6$ demonstrate. This is reflected in the drop of the mean radius. Subsequent growth of the inner radius brings it closer to the next node of the wave function what leads to the formation of the new $\overline{r}$ plateau being located above its predecessor since the function $f$ is distributed wider for the larger inner radius. The sharp drop of the mean radius after the last flat part is explained again by the accumulation of the wave function near the origin (see panel for $b=5$ in figure \ref{Fig7}). After this pronounced minimum the magnitude of $\overline{r}$ smoothly approaches its value of the two equal disc geometry which, quite naturally, coincides with the mean radius of the system with one Neumann window:
\begin{equation}
\left.\overline{r}\right|_{b=0}=\left.\overline{r}\right|_{b=a}.
\end{equation}
Similar to the energy, the transition between the $\overline{r}$ plateaus gets sharper for the larger outer radii, as the comparison of panels (e) and (f) of figure \ref{Fig6} demonstrates. 

Anticrossings very similar to the ones discussed above were calculated for the magnetic quantum ring \cite{Kim1} when the increasing inner radius of the field-free annulus in the otherwise uniform magnetic field for the large fixed values of its outer counterpart forms the avoided crossings of the energies of the adjacent states with the same azimuthal quantum number $m$. Passage through the energy anticrossing in this case is accompanied by the abrupt change of the corresponding whirling persistent current of the same quantum mechanical state what draws the very clear parallel to our geometry with the rapid variations of the mean radius $\overline{r}$. This similarity becomes almost complete if one recalls that the azimuthal current in the uniform magnetic field is determined by the expression from the right-hand side of equation \eref{MeanRho} \cite{Kim1}. Energy anticrossings present an ubiquitous feature  of the energy spectrum of the quantum systems with finite height potentials \cite{Kudryavtsev1,Maier1,Johnson1,Gesztesy1,Ancilotto1,Platero1,Falko1,Cruz1,Olendskii1,Ferreyra1,Lyo1,Ho1,Hagedorn1,Exner4,Olendski4,Arsoski1,Szafran1} and are indispensable, for example, in the description of the quantum Hall effect \cite{Ho1}.

Finally, we provide the proof of Proposition \ref{Proposition1}. For the large radius, the $z$ dependence of the total function $f(r,\theta,z)$ is defined by $\chi_j^{DD}(z)$, equation \eref{TransverseDD}. In turn, general solution of the radial equation \eref{bes} for $E=1$ and  $j=0$ is given as
\begin{equation}
R(r)=\left\{
\begin{array}{cc} C_1+C_2\ln r,&m=0\\
C_1r^{-|m|}+C_2r^{|m|},&m\ne 0
\end{array}
\right.
\end{equation}
with constants $C_1$ and $C_2$. From the demand of the vanishing of the function $f$ at infinity it immediately follows that $C_2\equiv 0$. In this way, the first term of the right-hand side of equation \eref{FunctionAsymptotics} is obtained. Remaining two terms there can be derived from \eref{SolutionOneWindow2} for $j=1$ and $j\ge 2$, respectively, and asymptotic properties of the Bessel functions $K_{|m|}(x)$ for the large arguments $x$ \cite{abra,wat}. We point out that the value of $C_1$ can be fixed by, say, the normalization condition, equation (\ref{normalization1}). This condition requires relatively fast decay of the wave function at infinity in order for the integral $\int_a^\infty\left|f_l^{(0)}\left(r,\theta,z\right)\right|^2rdr$ to be convergent. As is seen from equation (\ref{FunctionAsymptotics}), this can be achieved for the azimuthal numbers $|m|\geq 2$. For smaller $|m|$ the radial integral at $E=1$ diverges and, accordingly, the corresponding state is a resonance without square-integrable function $f_l^{(0)}$. The particular value of $C_1$ in this case can be determined say, from the $\delta$-normalization. We also note that the bound states and resonances at the critical radius, similar to the true bound states with $E<1$, are degenerate with respect to the sign of the azimuthal number $m$: the two states with the opposite signs of $m$ possess the same energy with their functions $f_l^{(0)}$ being complex conjugate ones.

\section{Conclusions}
\label{sec_5}
Our comparative analysis has confirmed that the two-Neumann-disc geometry of the otherwise uniform straight 3D Dirichlet wave guide, similar to its one-window counterpart, binds the Schr\"{o}dinger particle with its energy lying below the essential Dirichlet spectrum. In the former case the binding is stronger than in the latter configuration since the two discs present a larger perturbation to the particle motion in the channel what results in the smaller critical values of the discs radius at which a new bound state emerges from the continuum. Moreover, the change of the second disc diameter opens up an additional channel of manipulating of the spectral and localization properties of the structure and leads to new phenomena such as anticrossings in the energy-inner radius dependence. We have also drawn parallels to the 2D geometry and have revealed a drastic difference between the two cases; namely, if, for the quasi-one-dimensional situation, the state at the critical length with its energy equal to the Dirichlet fundamental propagation threshold is {\it always} a resonance with non vanishing at infinity wave function, in three dimensions this is true for the states with zero azimuthal quantum number, $m=0$. We have classified the state with $|m|=1$ as a resonance too since its function, even though vanishing at infinity, can not guarantee square integrability while for the larger $|m|$ one has true bound levels with square integrable function.

In our treatment we assumed that the transverse potential is flat. Applying the external electric field $\mbox{\boldmath$\cal E$}$ perpendicular to the interfaces will turn it into the linear tilted one. Influence of the transverse gate voltage $U={\cal E}/d$ on the electronic and optical properties of the pure Dirichlet quantum well has been the subject of intensive theoretical \cite{Rabinovitch1,Lukes1,Bastard1,Nguyen1,Miller1,Miller2} and experimental \cite{Miller1} research. It was shown that the field $\mbox{\boldmath$\cal E$}$ significantly changes charged particle behaviour in these low-dimensional nanostructures leading, for example, to the quantum-confined Stark effect \cite{Miller1}. One can expect that for our geometry, similar to the curved 2D Dirichlet wave guides \cite{Exner5}, the perpendicular electric field will have a noticeable influence on the bound states too. Such calculations are now in progress.

\section{Acknowledgements}
\label{sec_6}
H.N. acknowledges support from Research Unity 01/UR/15-01. O.O. acknowledges the warm hospitality of D\'epartemente de Math\'ematiques, ISMAI, Kairouan, Tunisia, where this project was initiated. Research at Jackson State University was financially supported in part by NSF under Grant No. DMR-0606509 and by DoD under contract No. W912HZ-06-C-0057. Stimulating discussions with Prof. P. Exner are gratefully appreciated.

\end{document}